%% file: main.tex
\documentclass[10pt]{article}

\usepackage{a4wide}
\usepackage{amsmath,amssymb}
\usepackage[colorlinks=true,bookmarks=false,citecolor=blue,urlcolor=blue]{hyperref} 
\usepackage{changes}
\usepackage{siunitx}
\usepackage{booktabs}
\usepackage{nicefrac}
\DeclareMathOperator*{\argmax}{argmax}
\usepackage[font=footnotesize]{caption}

\newcommand{\tcb}[1]{#1}
\usepackage{authblk}

\begin{document}

\title{Spiking Neural Network Equalization\\for IM/DD Optical Communication}
\author{Elias Arnold$^{1, \dagger}$, Georg B\"ocherer$^{2, \ddagger}$, Eric M\"uller$^{1}$, Philipp Spilger$^{1}$,\\Johannes Schemmel$^{1}$, Stefano Calabr\`o$^{2}$, Maxim Kuschnerov$^{2}$}
\affil{\scriptsize%
$^{1}$Electronic Vision(s), Kirchhoff-Institute for Physics, Heidelberg University, Germany\newline%
$^{2}$Huawei Technologies Duesseldorf GmbH, Munich Research Center, Germany}

\date{}

\maketitle

\begin{abstract}
	\input{abstract.tex}
\end{abstract}

\input{introduction}
\input{results}
\input{discussion}

\section*{Funding}
The contributions of the Electronic Vision(s) group\footnotemark[1] have been supported by
the EC Horizon 2020 Framework Programme under grant agreements
785907 (HBP SGA2)
and
945539 (HBP SGA3),
the Deutsche Forschungsgemeinschaft (DFG, German Research Foundation) under Germany's Excellence Strategy EXC 2181/1-390900948 (the Heidelberg STRUCTURES Excellence Cluster),
the Helmholtz Association Initiative and Networking Fund [Advanced Computing Architectures (ACA)] under Project SO-092.

\bibliography{bib}
\bibliographystyle{ieeetr}

\end{document}

%% file: abstract.tex
A spiking neural network (SNN) equalizer model suitable for electronic neuromorphic hardware is designed for an IM/DD link. The SNN achieves the same bit-error-rate as an artificial neural network, outperforming linear equalization.

%% file: introduction.tex
\section{Introduction}
Low cost and low power optical transceivers are indispensable for supporting the exponentially growing data center traffic caused by cloud-based services. The high power consumption of \emph{digital signal processing} (DSP) has motivated research on moving parts of the receiver DSP to an analog lower power frontend. For instance, photonic neuromorphic computing \cite{shastri2021photonics} has been studied recently, e.g., for compensating for \emph{chromatic dispersion} (CD) and nonlinear impairments in short reach optical transmission~\cite{li2021micro,ranzini2021experimental}. An alternative solution is 
analog electronic
neuromorphic computing, implementing SNNs~\cite{gerstner2014neuronal} in analog hardware~\cite{pehle2022brainscales2} by mimicking the basic operation principles of the human brain, thereby adopting the brain's unchallenged power efficiency. 
SNNs are applied in \cite{wu2022improved} for an inference task on a spectrogram in fiber-optic distributed acoustic sensing.
Recently, in-the-loop training of SNNs on analog hardware has achieved state-of-the-art performance in inference tasks \cite{cramer2022surrogate}. 
Despite electronics operating slower than photonics,
electronic hardware enables higher scalability and thus greater throughput through parallelization, making it a suitable choice for energy-efficient signal processing.
An important aspect to be analyzed is whether SNNs in 
analog electronic 
hardware support the accuracy required by communication systems.

To assess the accuracy of SNNs, we design equalization and demapping using SNNs suitable for a hardware implementation on the BrainScaleS-2 (BSS-2) system~\cite{pehle2022brainscales2}. We evaluate our SNN in a software simulation for the detection of a \emph{4-level pulse amplitude modulation} (PAM4) signal for an \emph{intensity modulation/direct detection} (IM/DD) link, impaired by CD 
and \emph{additive white Gaussian noise} (AWGN). Our SNN achieves the \emph{bit error rate} (BER) of an \emph{artificial neural network} (ANN), outperforming a digital \emph{linear minimum mean square error} (LMMSE) equalizer.
\begin{figure}[bp]
\footnotesize
\centering
\begin{tikzpicture}
\tikzset{
    neuron/.style={circle, inner sep=0pt, align=center, minimum size=9pt, draw},
    labelneuron/.style={circle, inner sep=1pt, align=center, minimum size=7pt, rounded corners=2pt},
    spike/.style={rectangle, fill=black!50, minimum width=1pt, minimum height=3pt, inner sep=0pt, draw=black!50},
    outer/.style={rectangle, align=center, draw=black, minimum height=2.3cm, minimum width=0.5cm, rounded corners=5pt},
    outer2/.style={rectangle, align=center, draw=black, minimum height=4.3cm, minimum width=0.5cm, rounded corners=5pt},
    outer3/.style={rectangle, align=center, draw=black, minimum height=1.8cm, minimum width=0.5cm, rounded corners=5pt}}
    
\def\dt{0.05}
\def\spacing{0.2}
\def\inoff{1.5}
\def\outoff{1.2}

\draw[-latex, very thick, draw=black] (-0.1, 4.4) -- (-0.1, 3.6);
\node () at (-0.3, 4.1) {$t$};
\node[rectangle, minimum height=2.2cm, minimum width=1.2cm, inner sep=4pt, align=center, rounded corners=5pt, draw=black, dotted] (input_otter) at (0.3, 0.9+\inoff) {};
\node[inner sep=4pt, align=center] (inputtm) at (0.3, 1.7+\inoff) {$y^{t-\lfloor \nicefrac{n_{\text{tap}}}{2}\rfloor}$};
\node[inner sep=4pt, align=center] (inputt) at (0.3, 0.9+\inoff) {$y^{t}$};
\node[inner sep=4pt, align=center] (inputtp) at (0.3, 0.1+\inoff) {$y^{t+\lfloor \nicefrac{n_{\text{tap}}}{2}\rfloor}$};
\node[inner sep=4pt, align=center] (inputt0) at (0.3, 2.8+\inoff) {$y^{0}$};

\draw[very thick, draw=black!50, dotted] (inputt0.south) -- (input_otter.north);
\draw[very thick, draw=black!50, dotted] (inputtm.south) -- (inputt.north);
\draw[very thick, draw=black!50, dotted] (inputt.south) -- (inputtp.north);

\node[outer] (inputs) at (1.6, 0.9+\inoff) {}; 
\node[neuron] (ni_0) at (1.6, \inoff) {};
\draw[very thick, dotted, draw=black!50] (1.6, 0.3+\inoff) -- (1.6, 1.5+\inoff);
\node[neuron] (ni_n) at (1.6, \inoff+1.8) {};

\node[inner sep=0pt] (traces) at (2.6, 2.6)
    {\input{./example_spikes.pgf}};
\draw[-latex, thick] (input_otter.east) -- (inputs.west);

\node[align=center] (label) at (2.4, 4.5) {spike encoding /\\input spike trains};
\node[align=center] (label) at (5.2, 0.0) {hidden LIF neurons};
\node[align=center] (label) at (8.8, 1.0) {output LI neurons};
\node[align=center] (eq) at (2.4, 0.5) {$\tau^s_i \sim  \log\left( \nicefrac{d}{(d - \beta)}\right)$\\with $d = |A - |y^t - \chi_i||$\\and $\beta, A$ const.};

\node[outer2] (hiddens) at (4.5, 0.9+\inoff) {};
\node[neuron] (n_0) at (4.5, 0.5) {};
\draw[very thick, dotted, draw=black!50] (4.5, 0.8) -- (4.5, 4.0);
\node[neuron] (n_1) at (4.5, 4.3) {};
\node[inner sep=0pt] (traces) at (5.5, 2.4)
    {\input{./example_spikes_hidden.pgf}};

\def\colors{{"1. 0. 0.", "0. 0. 1.", "1. 0.65 0.00", "0. .5 0",}}
\def\labels{{"0", "1", "2", "3"}}
\foreach \x in {0,...,3} {
    \pgfmathparse{\colors[\x]};
    \pgfmathtruncatemacro{\label}{\labels[\x]};
    \definecolor{color}{rgb}{\pgfmathresult};
    \node[labelneuron, draw=color] (no_\x) at (7.2, \outoff + 0.6 + 2.0*\spacing*\x) {\label};}
\node[outer3] (outputs) at (7.2, \outoff + 1.2) {};
\node[inner sep=0pt] (traces) at (9.1, 2.4)
    {\input{./example_traces.pgf}};

\draw [-latex, thick] (3.6, \inoff+0.9) -- (hiddens.west);
\node () at (3.9, \inoff+1.3) {$W^\text{ih}_{ji}$};

\draw [-latex, thick] (6.3, \inoff+0.9) -- (outputs.west);
\node () at (6.6, \inoff+1.3) {$W^\text{ho}_{kj}$};

\node[rectangle, inner sep=4pt, align=center, rounded corners=5pt, draw=black] (max) at (11.6, \outoff + 1.2) {$\max_{\tau}(v_k)$\\and\\$\argmax_k$};

\definecolor{green}{rgb}{0., .5, 0};
\node[rectangle, inner sep=4pt, rounded corners=5pt, draw=black, align=left] (map) at (13.3, \outoff + 1.2) {
\color{green}{$\mathbf{3\mapsto 10}$} \\
\color{gray}{$2\mapsto 11$}\\
\color{gray}{$1\mapsto 01$}\\
\color{gray}{$0 \mapsto 00$}};

\draw [-latex, thick] (10.5, \outoff + 1.2) -- (max.west);
\draw [-latex, thick] (max.east) -- (map.west);

\node[rectangle, inner sep=4pt, align=center, rounded corners=5pt, draw=black] (out) at (14.7, \outoff + 1.2) {$\hat{B}_1\hat{B}_2$};
\draw [-latex, thick] (map.east) -- (out.west);

\end{tikzpicture}
\vspace{-10pt}
\caption{SNN equalizer demapper decision chain}
\label{fig:demapper setup}
\end{figure}
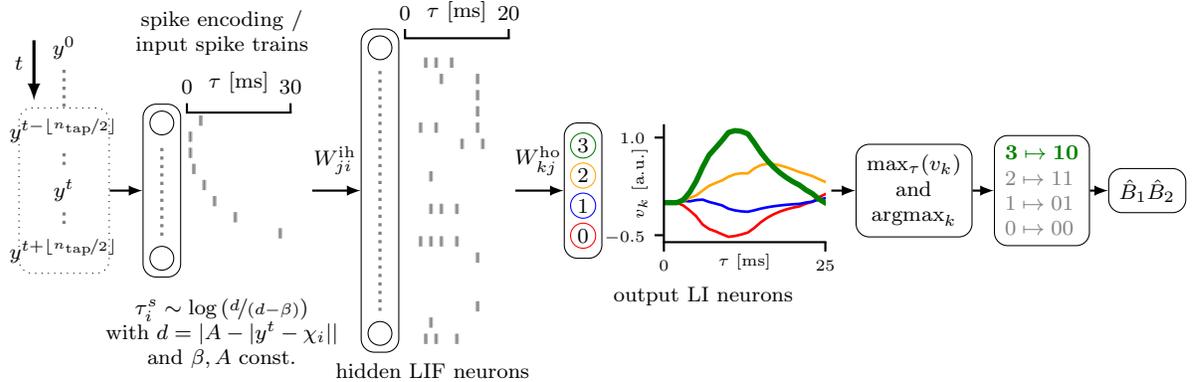

\section{Equalization and Demapping using Spiking Neural Networks}
For equalization and demapping, we consider an SNN with a single hidden layer, consisting of 40 spiking \emph{leaky-integrate and fire} (LIF) neurons~\cite[Sec.~1.3]{gerstner2014neuronal}, and an output layer constituted by four non-spiking \emph{leaky-integrate} (LI)~\cite[Sec.~1.3]{gerstner2014neuronal} readout neurons. This architecture fits in size on the BSS-2 system~\cite{pehle2022brainscales2}. Each LIF neuron $j$ maintains an internal membrane state $v_j$ described by the ordinary differential equations
\begin{align}
	\tau_\text{m} \dot{v}_j(\tau) = - \left(v_j(\tau) - v_\text{leak} \right) + I_j(\tau) \quad \text{with} \quad I_j(\tau) = \sum_{i=0}^{N-1} \sum_{s \in \lbrace \text{spikes }i\text{-th neuron}\rbrace} \notag\\ W_{ji} \Theta \left( \tau-\tau_i^s \right)\exp\left(-\frac{\tau-\tau^s_i}{\tau_\text{syn}}\right),
\end{align}
integrating synaptic input $I$, caused by pre-synaptic events $\tau_i^s$, onto its membrane. As the membrane potential exceeds a threshold $\vartheta$, the neuron emits a post-synaptic spike a time $\tau^s_j$, after which it is set to a reset potential $v_\text{r}$. LI neurons exhibit the same dynamics, without the ability to spike. The parameters $\tau_\text{syn}$ and $\tau_\text{m}$ are the time constants of the synaptic current and the membrane potential, respectively.

A received sample $y^t$ and its $\lfloor \nicefrac{n_{\text{tap}}}{2}\rfloor$ predecessors and successors ($n_\text{tap}$ odd) are translated to 10 input spike events per sample by a spike encoder (see Fig.~\ref{fig:demapper setup}),
potentially replacing power-hungry \emph{analog-to-digital conversion} (ADC) in hardware. To this end, each input neuron emits a spike at time $\tau^s_i$ given by the scaled log-distance~\cite{superspike} to a reference point $\chi_i$, assigned to each input neuron. The input sample $y^t$ gets classified with the label $k \in \lbrace 0, 1, 2, 3\rbrace$ of the output neuron with the maximum membrane value $v_k(\tau)$ over the considered time frame. Hence, the network learns to place hidden spike events in time, such that the readout traces are adjusted appropriately. 

For training our SNNs we rely on \emph{backpropagation through time} (BPTT) with the Adam optimizer and surrogate gradients (SuperSpike \cite{superspike}) to account for the discontinuity of spiking LIF neurons. Note that our simulations are implemented in \texttt{hxtorch}~\cite{mueller2022scalable}, also supporting execution on the BSS-2 system.

%% file: example_spikes.pgf
\begingroup%
\makeatletter%
\begin{pgfpicture}%
\pgfpathrectangle{\pgfpointorigin}{\pgfqpoint{0.700000in}{1.100000in}}%
\pgfusepath{use as bounding box, clip}%
\begin{pgfscope}%
\pgfsetbuttcap%
\pgfsetmiterjoin%
\pgfsetlinewidth{0.000000pt}%
\definecolor{currentstroke}{rgb}{0.000000,0.000000,0.000000}%
\pgfsetstrokecolor{currentstroke}%
\pgfsetstrokeopacity{0.000000}%
\pgfsetdash{}{0pt}%
\pgfpathmoveto{\pgfqpoint{0.000000in}{0.000000in}}%
\pgfpathlineto{\pgfqpoint{0.700000in}{0.000000in}}%
\pgfpathlineto{\pgfqpoint{0.700000in}{1.100000in}}%
\pgfpathlineto{\pgfqpoint{0.000000in}{1.100000in}}%
\pgfpathlineto{\pgfqpoint{0.000000in}{0.000000in}}%
\pgfpathclose%
\pgfusepath{}%
\end{pgfscope}%
\begin{pgfscope}%
\pgfsetbuttcap%
\pgfsetmiterjoin%
\pgfsetlinewidth{0.000000pt}%
\definecolor{currentstroke}{rgb}{0.000000,0.000000,0.000000}%
\pgfsetstrokecolor{currentstroke}%
\pgfsetstrokeopacity{0.000000}%
\pgfsetdash{}{0pt}%
\pgfpathmoveto{\pgfqpoint{0.087500in}{0.121000in}}%
\pgfpathlineto{\pgfqpoint{0.630000in}{0.121000in}}%
\pgfpathlineto{\pgfqpoint{0.630000in}{0.880000in}}%
\pgfpathlineto{\pgfqpoint{0.087500in}{0.880000in}}%
\pgfpathlineto{\pgfqpoint{0.087500in}{0.121000in}}%
\pgfpathclose%
\pgfusepath{}%
\end{pgfscope}%
\begin{pgfscope}%
\pgfpathrectangle{\pgfqpoint{0.087500in}{0.121000in}}{\pgfqpoint{0.542500in}{0.759000in}}%
\pgfusepath{clip}%
\pgfsetbuttcap%
\pgfsetroundjoin%
\definecolor{currentfill}{rgb}{0.184314,0.184314,0.184314}%
\pgfsetfillcolor{currentfill}%
\pgfsetfillopacity{0.400000}%
\pgfsetlinewidth{1.505625pt}%
\definecolor{currentstroke}{rgb}{0.184314,0.184314,0.184314}%
\pgfsetstrokecolor{currentstroke}%
\pgfsetstrokeopacity{0.400000}%
\pgfsetdash{}{0pt}%
\pgfsys@defobject{currentmarker}{\pgfqpoint{0.000000in}{-0.026896in}}{\pgfqpoint{0.000000in}{0.026896in}}{%
\pgfpathmoveto{\pgfqpoint{0.000000in}{-0.026896in}}%
\pgfpathlineto{\pgfqpoint{0.000000in}{0.026896in}}%
\pgfusepath{stroke,fill}%
}%
\end{pgfscope}%
\begin{pgfscope}%
\pgfpathrectangle{\pgfqpoint{0.087500in}{0.121000in}}{\pgfqpoint{0.542500in}{0.759000in}}%
\pgfusepath{clip}%
\pgfsetbuttcap%
\pgfsetroundjoin%
\definecolor{currentfill}{rgb}{0.184314,0.184314,0.184314}%
\pgfsetfillcolor{currentfill}%
\pgfsetfillopacity{0.400000}%
\pgfsetlinewidth{1.505625pt}%
\definecolor{currentstroke}{rgb}{0.184314,0.184314,0.184314}%
\pgfsetstrokecolor{currentstroke}%
\pgfsetstrokeopacity{0.400000}%
\pgfsetdash{}{0pt}%
\pgfsys@defobject{currentmarker}{\pgfqpoint{0.000000in}{-0.026896in}}{\pgfqpoint{0.000000in}{0.026896in}}{%
\pgfpathmoveto{\pgfqpoint{0.000000in}{-0.026896in}}%
\pgfpathlineto{\pgfqpoint{0.000000in}{0.026896in}}%
\pgfusepath{stroke,fill}%
}%
\end{pgfscope}%
\begin{pgfscope}%
\pgfpathrectangle{\pgfqpoint{0.087500in}{0.121000in}}{\pgfqpoint{0.542500in}{0.759000in}}%
\pgfusepath{clip}%
\pgfsetbuttcap%
\pgfsetroundjoin%
\definecolor{currentfill}{rgb}{0.184314,0.184314,0.184314}%
\pgfsetfillcolor{currentfill}%
\pgfsetfillopacity{0.400000}%
\pgfsetlinewidth{1.505625pt}%
\definecolor{currentstroke}{rgb}{0.184314,0.184314,0.184314}%
\pgfsetstrokecolor{currentstroke}%
\pgfsetstrokeopacity{0.400000}%
\pgfsetdash{}{0pt}%
\pgfsys@defobject{currentmarker}{\pgfqpoint{0.000000in}{-0.026896in}}{\pgfqpoint{0.000000in}{0.026896in}}{%
\pgfpathmoveto{\pgfqpoint{0.000000in}{-0.026896in}}%
\pgfpathlineto{\pgfqpoint{0.000000in}{0.026896in}}%
\pgfusepath{stroke,fill}%
}%
\begin{pgfscope}%
\pgfsys@transformshift{0.575750in}{0.247500in}%
\pgfsys@useobject{currentmarker}{}%
\end{pgfscope}%
\end{pgfscope}%
\begin{pgfscope}%
\pgfpathrectangle{\pgfqpoint{0.087500in}{0.121000in}}{\pgfqpoint{0.542500in}{0.759000in}}%
\pgfusepath{clip}%
\pgfsetbuttcap%
\pgfsetroundjoin%
\definecolor{currentfill}{rgb}{0.184314,0.184314,0.184314}%
\pgfsetfillcolor{currentfill}%
\pgfsetfillopacity{0.400000}%
\pgfsetlinewidth{1.505625pt}%
\definecolor{currentstroke}{rgb}{0.184314,0.184314,0.184314}%
\pgfsetstrokecolor{currentstroke}%
\pgfsetstrokeopacity{0.400000}%
\pgfsetdash{}{0pt}%
\pgfsys@defobject{currentmarker}{\pgfqpoint{0.000000in}{-0.026896in}}{\pgfqpoint{0.000000in}{0.026896in}}{%
\pgfpathmoveto{\pgfqpoint{0.000000in}{-0.026896in}}%
\pgfpathlineto{\pgfqpoint{0.000000in}{0.026896in}}%
\pgfusepath{stroke,fill}%
}%
\begin{pgfscope}%
\pgfsys@transformshift{0.340667in}{0.331833in}%
\pgfsys@useobject{currentmarker}{}%
\end{pgfscope}%
\end{pgfscope}%
\begin{pgfscope}%
\pgfpathrectangle{\pgfqpoint{0.087500in}{0.121000in}}{\pgfqpoint{0.542500in}{0.759000in}}%
\pgfusepath{clip}%
\pgfsetbuttcap%
\pgfsetroundjoin%
\definecolor{currentfill}{rgb}{0.184314,0.184314,0.184314}%
\pgfsetfillcolor{currentfill}%
\pgfsetfillopacity{0.400000}%
\pgfsetlinewidth{1.505625pt}%
\definecolor{currentstroke}{rgb}{0.184314,0.184314,0.184314}%
\pgfsetstrokecolor{currentstroke}%
\pgfsetstrokeopacity{0.400000}%
\pgfsetdash{}{0pt}%
\pgfsys@defobject{currentmarker}{\pgfqpoint{0.000000in}{-0.026896in}}{\pgfqpoint{0.000000in}{0.026896in}}{%
\pgfpathmoveto{\pgfqpoint{0.000000in}{-0.026896in}}%
\pgfpathlineto{\pgfqpoint{0.000000in}{0.026896in}}%
\pgfusepath{stroke,fill}%
}%
\begin{pgfscope}%
\pgfsys@transformshift{0.232167in}{0.416167in}%
\pgfsys@useobject{currentmarker}{}%
\end{pgfscope}%
\end{pgfscope}%
\begin{pgfscope}%
\pgfpathrectangle{\pgfqpoint{0.087500in}{0.121000in}}{\pgfqpoint{0.542500in}{0.759000in}}%
\pgfusepath{clip}%
\pgfsetbuttcap%
\pgfsetroundjoin%
\definecolor{currentfill}{rgb}{0.184314,0.184314,0.184314}%
\pgfsetfillcolor{currentfill}%
\pgfsetfillopacity{0.400000}%
\pgfsetlinewidth{1.505625pt}%
\definecolor{currentstroke}{rgb}{0.184314,0.184314,0.184314}%
\pgfsetstrokecolor{currentstroke}%
\pgfsetstrokeopacity{0.400000}%
\pgfsetdash{}{0pt}%
\pgfsys@defobject{currentmarker}{\pgfqpoint{0.000000in}{-0.026896in}}{\pgfqpoint{0.000000in}{0.026896in}}{%
\pgfpathmoveto{\pgfqpoint{0.000000in}{-0.026896in}}%
\pgfpathlineto{\pgfqpoint{0.000000in}{0.026896in}}%
\pgfusepath{stroke,fill}%
}%
\begin{pgfscope}%
\pgfsys@transformshift{0.177917in}{0.500500in}%
\pgfsys@useobject{currentmarker}{}%
\end{pgfscope}%
\end{pgfscope}%
\begin{pgfscope}%
\pgfpathrectangle{\pgfqpoint{0.087500in}{0.121000in}}{\pgfqpoint{0.542500in}{0.759000in}}%
\pgfusepath{clip}%
\pgfsetbuttcap%
\pgfsetroundjoin%
\definecolor{currentfill}{rgb}{0.184314,0.184314,0.184314}%
\pgfsetfillcolor{currentfill}%
\pgfsetfillopacity{0.400000}%
\pgfsetlinewidth{1.505625pt}%
\definecolor{currentstroke}{rgb}{0.184314,0.184314,0.184314}%
\pgfsetstrokecolor{currentstroke}%
\pgfsetstrokeopacity{0.400000}%
\pgfsetdash{}{0pt}%
\pgfsys@defobject{currentmarker}{\pgfqpoint{0.000000in}{-0.026896in}}{\pgfqpoint{0.000000in}{0.026896in}}{%
\pgfpathmoveto{\pgfqpoint{0.000000in}{-0.026896in}}%
\pgfpathlineto{\pgfqpoint{0.000000in}{0.026896in}}%
\pgfusepath{stroke,fill}%
}%
\begin{pgfscope}%
\pgfsys@transformshift{0.123667in}{0.584833in}%
\pgfsys@useobject{currentmarker}{}%
\end{pgfscope}%
\end{pgfscope}%
\begin{pgfscope}%
\pgfpathrectangle{\pgfqpoint{0.087500in}{0.121000in}}{\pgfqpoint{0.542500in}{0.759000in}}%
\pgfusepath{clip}%
\pgfsetbuttcap%
\pgfsetroundjoin%
\definecolor{currentfill}{rgb}{0.184314,0.184314,0.184314}%
\pgfsetfillcolor{currentfill}%
\pgfsetfillopacity{0.400000}%
\pgfsetlinewidth{1.505625pt}%
\definecolor{currentstroke}{rgb}{0.184314,0.184314,0.184314}%
\pgfsetstrokecolor{currentstroke}%
\pgfsetstrokeopacity{0.400000}%
\pgfsetdash{}{0pt}%
\pgfsys@defobject{currentmarker}{\pgfqpoint{0.000000in}{-0.026896in}}{\pgfqpoint{0.000000in}{0.026896in}}{%
\pgfpathmoveto{\pgfqpoint{0.000000in}{-0.026896in}}%
\pgfpathlineto{\pgfqpoint{0.000000in}{0.026896in}}%
\pgfusepath{stroke,fill}%
}%
\begin{pgfscope}%
\pgfsys@transformshift{0.105583in}{0.669167in}%
\pgfsys@useobject{currentmarker}{}%
\end{pgfscope}%
\end{pgfscope}%
\begin{pgfscope}%
\pgfpathrectangle{\pgfqpoint{0.087500in}{0.121000in}}{\pgfqpoint{0.542500in}{0.759000in}}%
\pgfusepath{clip}%
\pgfsetbuttcap%
\pgfsetroundjoin%
\definecolor{currentfill}{rgb}{0.184314,0.184314,0.184314}%
\pgfsetfillcolor{currentfill}%
\pgfsetfillopacity{0.400000}%
\pgfsetlinewidth{1.505625pt}%
\definecolor{currentstroke}{rgb}{0.184314,0.184314,0.184314}%
\pgfsetstrokecolor{currentstroke}%
\pgfsetstrokeopacity{0.400000}%
\pgfsetdash{}{0pt}%
\pgfsys@defobject{currentmarker}{\pgfqpoint{0.000000in}{-0.026896in}}{\pgfqpoint{0.000000in}{0.026896in}}{%
\pgfpathmoveto{\pgfqpoint{0.000000in}{-0.026896in}}%
\pgfpathlineto{\pgfqpoint{0.000000in}{0.026896in}}%
\pgfusepath{stroke,fill}%
}%
\begin{pgfscope}%
\pgfsys@transformshift{0.105583in}{0.753500in}%
\pgfsys@useobject{currentmarker}{}%
\end{pgfscope}%
\end{pgfscope}%
\begin{pgfscope}%
\pgfpathrectangle{\pgfqpoint{0.087500in}{0.121000in}}{\pgfqpoint{0.542500in}{0.759000in}}%
\pgfusepath{clip}%
\pgfsetbuttcap%
\pgfsetroundjoin%
\definecolor{currentfill}{rgb}{0.184314,0.184314,0.184314}%
\pgfsetfillcolor{currentfill}%
\pgfsetfillopacity{0.400000}%
\pgfsetlinewidth{1.505625pt}%
\definecolor{currentstroke}{rgb}{0.184314,0.184314,0.184314}%
\pgfsetstrokecolor{currentstroke}%
\pgfsetstrokeopacity{0.400000}%
\pgfsetdash{}{0pt}%
\pgfsys@defobject{currentmarker}{\pgfqpoint{0.000000in}{-0.026896in}}{\pgfqpoint{0.000000in}{0.026896in}}{%
\pgfpathmoveto{\pgfqpoint{0.000000in}{-0.026896in}}%
\pgfpathlineto{\pgfqpoint{0.000000in}{0.026896in}}%
\pgfusepath{stroke,fill}%
}%
\begin{pgfscope}%
\pgfsys@transformshift{0.159833in}{0.837833in}%
\pgfsys@useobject{currentmarker}{}%
\end{pgfscope}%
\end{pgfscope}%
\begin{pgfscope}%
\pgfsetbuttcap%
\pgfsetroundjoin%
\definecolor{currentfill}{rgb}{0.000000,0.000000,0.000000}%
\pgfsetfillcolor{currentfill}%
\pgfsetlinewidth{0.803000pt}%
\definecolor{currentstroke}{rgb}{0.000000,0.000000,0.000000}%
\pgfsetstrokecolor{currentstroke}%
\pgfsetdash{}{0pt}%
\pgfsys@defobject{currentmarker}{\pgfqpoint{0.000000in}{0.000000in}}{\pgfqpoint{0.000000in}{0.048611in}}{%
\pgfpathmoveto{\pgfqpoint{0.000000in}{0.000000in}}%
\pgfpathlineto{\pgfqpoint{0.000000in}{0.048611in}}%
\pgfusepath{stroke,fill}%
}%
\begin{pgfscope}%
\pgfsys@transformshift{0.087500in}{0.880000in}%
\pgfsys@useobject{currentmarker}{}%
\end{pgfscope}%
\end{pgfscope}%
\begin{pgfscope}%
\definecolor{textcolor}{rgb}{0.000000,0.000000,0.000000}%
\pgfsetstrokecolor{textcolor}%
\pgfsetfillcolor{textcolor}%
\pgftext[x=0.087500in,y=0.977222in,,bottom]{\color{textcolor}\rmfamily\fontsize{8.000000}{9.600000}\selectfont 0}%
\end{pgfscope}%
\begin{pgfscope}%
\pgfsetbuttcap%
\pgfsetroundjoin%
\definecolor{currentfill}{rgb}{0.000000,0.000000,0.000000}%
\pgfsetfillcolor{currentfill}%
\pgfsetlinewidth{0.803000pt}%
\definecolor{currentstroke}{rgb}{0.000000,0.000000,0.000000}%
\pgfsetstrokecolor{currentstroke}%
\pgfsetdash{}{0pt}%
\pgfsys@defobject{currentmarker}{\pgfqpoint{0.000000in}{0.000000in}}{\pgfqpoint{0.000000in}{0.048611in}}{%
\pgfpathmoveto{\pgfqpoint{0.000000in}{0.000000in}}%
\pgfpathlineto{\pgfqpoint{0.000000in}{0.048611in}}%
\pgfusepath{stroke,fill}%
}%
\begin{pgfscope}%
\pgfsys@transformshift{0.630000in}{0.880000in}%
\pgfsys@useobject{currentmarker}{}%
\end{pgfscope}%
\end{pgfscope}%
\begin{pgfscope}%
\definecolor{textcolor}{rgb}{0.000000,0.000000,0.000000}%
\pgfsetstrokecolor{textcolor}%
\pgfsetfillcolor{textcolor}%
\pgftext[x=0.630000in,y=0.977222in,,bottom]{\color{textcolor}\rmfamily\fontsize{8.000000}{9.600000}\selectfont 30}%
\end{pgfscope}%
\begin{pgfscope}%
\definecolor{textcolor}{rgb}{0.000000,0.000000,0.000000}%
\pgfsetstrokecolor{textcolor}%
\pgfsetfillcolor{textcolor}%
\pgftext[x=0.358750in,y=1.009030in,,base]{\color{textcolor}\rmfamily\fontsize{8.000000}{9.600000}\selectfont \(\displaystyle \tau\) [ms]}%
\end{pgfscope}%
\begin{pgfscope}%
\pgfsetrectcap%
\pgfsetmiterjoin%
\pgfsetlinewidth{0.803000pt}%
\definecolor{currentstroke}{rgb}{0.000000,0.000000,0.000000}%
\pgfsetstrokecolor{currentstroke}%
\pgfsetdash{}{0pt}%
\pgfpathmoveto{\pgfqpoint{0.087500in}{0.880000in}}%
\pgfpathlineto{\pgfqpoint{0.630000in}{0.880000in}}%
\pgfusepath{stroke}%
\end{pgfscope}%
\end{pgfpicture}%
\makeatother%
\endgroup%

%% file: example_spikes_hidden.pgf
\begingroup%
\makeatletter%
\begin{pgfpicture}%
\pgfpathrectangle{\pgfpointorigin}{\pgfqpoint{0.700000in}{2.100000in}}%
\pgfusepath{use as bounding box, clip}%
\begin{pgfscope}%
\pgfsetbuttcap%
\pgfsetmiterjoin%
\pgfsetlinewidth{0.000000pt}%
\definecolor{currentstroke}{rgb}{0.000000,0.000000,0.000000}%
\pgfsetstrokecolor{currentstroke}%
\pgfsetstrokeopacity{0.000000}%
\pgfsetdash{}{0pt}%
\pgfpathmoveto{\pgfqpoint{0.000000in}{0.000000in}}%
\pgfpathlineto{\pgfqpoint{0.700000in}{0.000000in}}%
\pgfpathlineto{\pgfqpoint{0.700000in}{2.100000in}}%
\pgfpathlineto{\pgfqpoint{0.000000in}{2.100000in}}%
\pgfpathlineto{\pgfqpoint{0.000000in}{0.000000in}}%
\pgfpathclose%
\pgfusepath{}%
\end{pgfscope}%
\begin{pgfscope}%
\pgfsetbuttcap%
\pgfsetmiterjoin%
\pgfsetlinewidth{0.000000pt}%
\definecolor{currentstroke}{rgb}{0.000000,0.000000,0.000000}%
\pgfsetstrokecolor{currentstroke}%
\pgfsetstrokeopacity{0.000000}%
\pgfsetdash{}{0pt}%
\pgfpathmoveto{\pgfqpoint{0.087500in}{0.231000in}}%
\pgfpathlineto{\pgfqpoint{0.630000in}{0.231000in}}%
\pgfpathlineto{\pgfqpoint{0.630000in}{1.848000in}}%
\pgfpathlineto{\pgfqpoint{0.087500in}{1.848000in}}%
\pgfpathlineto{\pgfqpoint{0.087500in}{0.231000in}}%
\pgfpathclose%
\pgfusepath{}%
\end{pgfscope}%
\begin{pgfscope}%
\pgfpathrectangle{\pgfqpoint{0.087500in}{0.231000in}}{\pgfqpoint{0.542500in}{1.617000in}}%
\pgfusepath{clip}%
\pgfsetbuttcap%
\pgfsetroundjoin%
\definecolor{currentfill}{rgb}{0.184314,0.184314,0.184314}%
\pgfsetfillcolor{currentfill}%
\pgfsetfillopacity{0.400000}%
\pgfsetlinewidth{1.505625pt}%
\definecolor{currentstroke}{rgb}{0.184314,0.184314,0.184314}%
\pgfsetstrokecolor{currentstroke}%
\pgfsetstrokeopacity{0.400000}%
\pgfsetdash{}{0pt}%
\pgfsys@defobject{currentmarker}{\pgfqpoint{0.000000in}{-0.026896in}}{\pgfqpoint{0.000000in}{0.026896in}}{%
\pgfpathmoveto{\pgfqpoint{0.000000in}{-0.026896in}}%
\pgfpathlineto{\pgfqpoint{0.000000in}{0.026896in}}%
\pgfusepath{stroke,fill}%
}%
\begin{pgfscope}%
\pgfsys@transformshift{0.196000in}{0.188447in}%
\pgfsys@useobject{currentmarker}{}%
\end{pgfscope}%
\begin{pgfscope}%
\pgfsys@transformshift{0.250250in}{0.188447in}%
\pgfsys@useobject{currentmarker}{}%
\end{pgfscope}%
\begin{pgfscope}%
\pgfsys@transformshift{0.331625in}{0.188447in}%
\pgfsys@useobject{currentmarker}{}%
\end{pgfscope}%
\end{pgfscope}%
\begin{pgfscope}%
\pgfpathrectangle{\pgfqpoint{0.087500in}{0.231000in}}{\pgfqpoint{0.542500in}{1.617000in}}%
\pgfusepath{clip}%
\pgfsetbuttcap%
\pgfsetroundjoin%
\definecolor{currentfill}{rgb}{0.184314,0.184314,0.184314}%
\pgfsetfillcolor{currentfill}%
\pgfsetfillopacity{0.400000}%
\pgfsetlinewidth{1.505625pt}%
\definecolor{currentstroke}{rgb}{0.184314,0.184314,0.184314}%
\pgfsetstrokecolor{currentstroke}%
\pgfsetstrokeopacity{0.400000}%
\pgfsetdash{}{0pt}%
\pgfsys@defobject{currentmarker}{\pgfqpoint{0.000000in}{-0.026896in}}{\pgfqpoint{0.000000in}{0.026896in}}{%
\pgfpathmoveto{\pgfqpoint{0.000000in}{-0.026896in}}%
\pgfpathlineto{\pgfqpoint{0.000000in}{0.026896in}}%
\pgfusepath{stroke,fill}%
}%
\begin{pgfscope}%
\pgfsys@transformshift{0.196000in}{0.273553in}%
\pgfsys@useobject{currentmarker}{}%
\end{pgfscope}%
\begin{pgfscope}%
\pgfsys@transformshift{0.250250in}{0.273553in}%
\pgfsys@useobject{currentmarker}{}%
\end{pgfscope}%
\begin{pgfscope}%
\pgfsys@transformshift{0.358750in}{0.273553in}%
\pgfsys@useobject{currentmarker}{}%
\end{pgfscope}%
\end{pgfscope}%
\begin{pgfscope}%
\pgfpathrectangle{\pgfqpoint{0.087500in}{0.231000in}}{\pgfqpoint{0.542500in}{1.617000in}}%
\pgfusepath{clip}%
\pgfsetbuttcap%
\pgfsetroundjoin%
\definecolor{currentfill}{rgb}{0.184314,0.184314,0.184314}%
\pgfsetfillcolor{currentfill}%
\pgfsetfillopacity{0.400000}%
\pgfsetlinewidth{1.505625pt}%
\definecolor{currentstroke}{rgb}{0.184314,0.184314,0.184314}%
\pgfsetstrokecolor{currentstroke}%
\pgfsetstrokeopacity{0.400000}%
\pgfsetdash{}{0pt}%
\pgfsys@defobject{currentmarker}{\pgfqpoint{0.000000in}{-0.026896in}}{\pgfqpoint{0.000000in}{0.026896in}}{%
\pgfpathmoveto{\pgfqpoint{0.000000in}{-0.026896in}}%
\pgfpathlineto{\pgfqpoint{0.000000in}{0.026896in}}%
\pgfusepath{stroke,fill}%
}%
\begin{pgfscope}%
\pgfsys@transformshift{0.223125in}{0.358658in}%
\pgfsys@useobject{currentmarker}{}%
\end{pgfscope}%
\end{pgfscope}%
\begin{pgfscope}%
\pgfpathrectangle{\pgfqpoint{0.087500in}{0.231000in}}{\pgfqpoint{0.542500in}{1.617000in}}%
\pgfusepath{clip}%
\pgfsetbuttcap%
\pgfsetroundjoin%
\definecolor{currentfill}{rgb}{0.184314,0.184314,0.184314}%
\pgfsetfillcolor{currentfill}%
\pgfsetfillopacity{0.400000}%
\pgfsetlinewidth{1.505625pt}%
\definecolor{currentstroke}{rgb}{0.184314,0.184314,0.184314}%
\pgfsetstrokecolor{currentstroke}%
\pgfsetstrokeopacity{0.400000}%
\pgfsetdash{}{0pt}%
\pgfsys@defobject{currentmarker}{\pgfqpoint{0.000000in}{-0.026896in}}{\pgfqpoint{0.000000in}{0.026896in}}{%
\pgfpathmoveto{\pgfqpoint{0.000000in}{-0.026896in}}%
\pgfpathlineto{\pgfqpoint{0.000000in}{0.026896in}}%
\pgfusepath{stroke,fill}%
}%
\begin{pgfscope}%
\pgfsys@transformshift{0.467250in}{0.443763in}%
\pgfsys@useobject{currentmarker}{}%
\end{pgfscope}%
\end{pgfscope}%
\begin{pgfscope}%
\pgfpathrectangle{\pgfqpoint{0.087500in}{0.231000in}}{\pgfqpoint{0.542500in}{1.617000in}}%
\pgfusepath{clip}%
\pgfsetbuttcap%
\pgfsetroundjoin%
\definecolor{currentfill}{rgb}{0.184314,0.184314,0.184314}%
\pgfsetfillcolor{currentfill}%
\pgfsetfillopacity{0.400000}%
\pgfsetlinewidth{1.505625pt}%
\definecolor{currentstroke}{rgb}{0.184314,0.184314,0.184314}%
\pgfsetstrokecolor{currentstroke}%
\pgfsetstrokeopacity{0.400000}%
\pgfsetdash{}{0pt}%
\pgfsys@defobject{currentmarker}{\pgfqpoint{0.000000in}{-0.026896in}}{\pgfqpoint{0.000000in}{0.026896in}}{%
\pgfpathmoveto{\pgfqpoint{0.000000in}{-0.026896in}}%
\pgfpathlineto{\pgfqpoint{0.000000in}{0.026896in}}%
\pgfusepath{stroke,fill}%
}%
\end{pgfscope}%
\begin{pgfscope}%
\pgfpathrectangle{\pgfqpoint{0.087500in}{0.231000in}}{\pgfqpoint{0.542500in}{1.617000in}}%
\pgfusepath{clip}%
\pgfsetbuttcap%
\pgfsetroundjoin%
\definecolor{currentfill}{rgb}{0.184314,0.184314,0.184314}%
\pgfsetfillcolor{currentfill}%
\pgfsetfillopacity{0.400000}%
\pgfsetlinewidth{1.505625pt}%
\definecolor{currentstroke}{rgb}{0.184314,0.184314,0.184314}%
\pgfsetstrokecolor{currentstroke}%
\pgfsetstrokeopacity{0.400000}%
\pgfsetdash{}{0pt}%
\pgfsys@defobject{currentmarker}{\pgfqpoint{0.000000in}{-0.026896in}}{\pgfqpoint{0.000000in}{0.026896in}}{%
\pgfpathmoveto{\pgfqpoint{0.000000in}{-0.026896in}}%
\pgfpathlineto{\pgfqpoint{0.000000in}{0.026896in}}%
\pgfusepath{stroke,fill}%
}%
\end{pgfscope}%
\begin{pgfscope}%
\pgfpathrectangle{\pgfqpoint{0.087500in}{0.231000in}}{\pgfqpoint{0.542500in}{1.617000in}}%
\pgfusepath{clip}%
\pgfsetbuttcap%
\pgfsetroundjoin%
\definecolor{currentfill}{rgb}{0.184314,0.184314,0.184314}%
\pgfsetfillcolor{currentfill}%
\pgfsetfillopacity{0.400000}%
\pgfsetlinewidth{1.505625pt}%
\definecolor{currentstroke}{rgb}{0.184314,0.184314,0.184314}%
\pgfsetstrokecolor{currentstroke}%
\pgfsetstrokeopacity{0.400000}%
\pgfsetdash{}{0pt}%
\pgfsys@defobject{currentmarker}{\pgfqpoint{0.000000in}{-0.026896in}}{\pgfqpoint{0.000000in}{0.026896in}}{%
\pgfpathmoveto{\pgfqpoint{0.000000in}{-0.026896in}}%
\pgfpathlineto{\pgfqpoint{0.000000in}{0.026896in}}%
\pgfusepath{stroke,fill}%
}%
\begin{pgfscope}%
\pgfsys@transformshift{0.467250in}{0.699079in}%
\pgfsys@useobject{currentmarker}{}%
\end{pgfscope}%
\end{pgfscope}%
\begin{pgfscope}%
\pgfpathrectangle{\pgfqpoint{0.087500in}{0.231000in}}{\pgfqpoint{0.542500in}{1.617000in}}%
\pgfusepath{clip}%
\pgfsetbuttcap%
\pgfsetroundjoin%
\definecolor{currentfill}{rgb}{0.184314,0.184314,0.184314}%
\pgfsetfillcolor{currentfill}%
\pgfsetfillopacity{0.400000}%
\pgfsetlinewidth{1.505625pt}%
\definecolor{currentstroke}{rgb}{0.184314,0.184314,0.184314}%
\pgfsetstrokecolor{currentstroke}%
\pgfsetstrokeopacity{0.400000}%
\pgfsetdash{}{0pt}%
\pgfsys@defobject{currentmarker}{\pgfqpoint{0.000000in}{-0.026896in}}{\pgfqpoint{0.000000in}{0.026896in}}{%
\pgfpathmoveto{\pgfqpoint{0.000000in}{-0.026896in}}%
\pgfpathlineto{\pgfqpoint{0.000000in}{0.026896in}}%
\pgfusepath{stroke,fill}%
}%
\begin{pgfscope}%
\pgfsys@transformshift{0.168875in}{0.784184in}%
\pgfsys@useobject{currentmarker}{}%
\end{pgfscope}%
\begin{pgfscope}%
\pgfsys@transformshift{0.223125in}{0.784184in}%
\pgfsys@useobject{currentmarker}{}%
\end{pgfscope}%
\begin{pgfscope}%
\pgfsys@transformshift{0.277375in}{0.784184in}%
\pgfsys@useobject{currentmarker}{}%
\end{pgfscope}%
\begin{pgfscope}%
\pgfsys@transformshift{0.358750in}{0.784184in}%
\pgfsys@useobject{currentmarker}{}%
\end{pgfscope}%
\end{pgfscope}%
\begin{pgfscope}%
\pgfpathrectangle{\pgfqpoint{0.087500in}{0.231000in}}{\pgfqpoint{0.542500in}{1.617000in}}%
\pgfusepath{clip}%
\pgfsetbuttcap%
\pgfsetroundjoin%
\definecolor{currentfill}{rgb}{0.184314,0.184314,0.184314}%
\pgfsetfillcolor{currentfill}%
\pgfsetfillopacity{0.400000}%
\pgfsetlinewidth{1.505625pt}%
\definecolor{currentstroke}{rgb}{0.184314,0.184314,0.184314}%
\pgfsetstrokecolor{currentstroke}%
\pgfsetstrokeopacity{0.400000}%
\pgfsetdash{}{0pt}%
\pgfsys@defobject{currentmarker}{\pgfqpoint{0.000000in}{-0.026896in}}{\pgfqpoint{0.000000in}{0.026896in}}{%
\pgfpathmoveto{\pgfqpoint{0.000000in}{-0.026896in}}%
\pgfpathlineto{\pgfqpoint{0.000000in}{0.026896in}}%
\pgfusepath{stroke,fill}%
}%
\end{pgfscope}%
\begin{pgfscope}%
\pgfpathrectangle{\pgfqpoint{0.087500in}{0.231000in}}{\pgfqpoint{0.542500in}{1.617000in}}%
\pgfusepath{clip}%
\pgfsetbuttcap%
\pgfsetroundjoin%
\definecolor{currentfill}{rgb}{0.184314,0.184314,0.184314}%
\pgfsetfillcolor{currentfill}%
\pgfsetfillopacity{0.400000}%
\pgfsetlinewidth{1.505625pt}%
\definecolor{currentstroke}{rgb}{0.184314,0.184314,0.184314}%
\pgfsetstrokecolor{currentstroke}%
\pgfsetstrokeopacity{0.400000}%
\pgfsetdash{}{0pt}%
\pgfsys@defobject{currentmarker}{\pgfqpoint{0.000000in}{-0.026896in}}{\pgfqpoint{0.000000in}{0.026896in}}{%
\pgfpathmoveto{\pgfqpoint{0.000000in}{-0.026896in}}%
\pgfpathlineto{\pgfqpoint{0.000000in}{0.026896in}}%
\pgfusepath{stroke,fill}%
}%
\begin{pgfscope}%
\pgfsys@transformshift{0.223125in}{0.954395in}%
\pgfsys@useobject{currentmarker}{}%
\end{pgfscope}%
\begin{pgfscope}%
\pgfsys@transformshift{0.277375in}{0.954395in}%
\pgfsys@useobject{currentmarker}{}%
\end{pgfscope}%
\begin{pgfscope}%
\pgfsys@transformshift{0.358750in}{0.954395in}%
\pgfsys@useobject{currentmarker}{}%
\end{pgfscope}%
\end{pgfscope}%
\begin{pgfscope}%
\pgfpathrectangle{\pgfqpoint{0.087500in}{0.231000in}}{\pgfqpoint{0.542500in}{1.617000in}}%
\pgfusepath{clip}%
\pgfsetbuttcap%
\pgfsetroundjoin%
\definecolor{currentfill}{rgb}{0.184314,0.184314,0.184314}%
\pgfsetfillcolor{currentfill}%
\pgfsetfillopacity{0.400000}%
\pgfsetlinewidth{1.505625pt}%
\definecolor{currentstroke}{rgb}{0.184314,0.184314,0.184314}%
\pgfsetstrokecolor{currentstroke}%
\pgfsetstrokeopacity{0.400000}%
\pgfsetdash{}{0pt}%
\pgfsys@defobject{currentmarker}{\pgfqpoint{0.000000in}{-0.026896in}}{\pgfqpoint{0.000000in}{0.026896in}}{%
\pgfpathmoveto{\pgfqpoint{0.000000in}{-0.026896in}}%
\pgfpathlineto{\pgfqpoint{0.000000in}{0.026896in}}%
\pgfusepath{stroke,fill}%
}%
\end{pgfscope}%
\begin{pgfscope}%
\pgfpathrectangle{\pgfqpoint{0.087500in}{0.231000in}}{\pgfqpoint{0.542500in}{1.617000in}}%
\pgfusepath{clip}%
\pgfsetbuttcap%
\pgfsetroundjoin%
\definecolor{currentfill}{rgb}{0.184314,0.184314,0.184314}%
\pgfsetfillcolor{currentfill}%
\pgfsetfillopacity{0.400000}%
\pgfsetlinewidth{1.505625pt}%
\definecolor{currentstroke}{rgb}{0.184314,0.184314,0.184314}%
\pgfsetstrokecolor{currentstroke}%
\pgfsetstrokeopacity{0.400000}%
\pgfsetdash{}{0pt}%
\pgfsys@defobject{currentmarker}{\pgfqpoint{0.000000in}{-0.026896in}}{\pgfqpoint{0.000000in}{0.026896in}}{%
\pgfpathmoveto{\pgfqpoint{0.000000in}{-0.026896in}}%
\pgfpathlineto{\pgfqpoint{0.000000in}{0.026896in}}%
\pgfusepath{stroke,fill}%
}%
\begin{pgfscope}%
\pgfsys@transformshift{0.223125in}{1.124605in}%
\pgfsys@useobject{currentmarker}{}%
\end{pgfscope}%
\end{pgfscope}%
\begin{pgfscope}%
\pgfpathrectangle{\pgfqpoint{0.087500in}{0.231000in}}{\pgfqpoint{0.542500in}{1.617000in}}%
\pgfusepath{clip}%
\pgfsetbuttcap%
\pgfsetroundjoin%
\definecolor{currentfill}{rgb}{0.184314,0.184314,0.184314}%
\pgfsetfillcolor{currentfill}%
\pgfsetfillopacity{0.400000}%
\pgfsetlinewidth{1.505625pt}%
\definecolor{currentstroke}{rgb}{0.184314,0.184314,0.184314}%
\pgfsetstrokecolor{currentstroke}%
\pgfsetstrokeopacity{0.400000}%
\pgfsetdash{}{0pt}%
\pgfsys@defobject{currentmarker}{\pgfqpoint{0.000000in}{-0.026896in}}{\pgfqpoint{0.000000in}{0.026896in}}{%
\pgfpathmoveto{\pgfqpoint{0.000000in}{-0.026896in}}%
\pgfpathlineto{\pgfqpoint{0.000000in}{0.026896in}}%
\pgfusepath{stroke,fill}%
}%
\begin{pgfscope}%
\pgfsys@transformshift{0.738500in}{1.209711in}%
\pgfsys@useobject{currentmarker}{}%
\end{pgfscope}%
\end{pgfscope}%
\begin{pgfscope}%
\pgfpathrectangle{\pgfqpoint{0.087500in}{0.231000in}}{\pgfqpoint{0.542500in}{1.617000in}}%
\pgfusepath{clip}%
\pgfsetbuttcap%
\pgfsetroundjoin%
\definecolor{currentfill}{rgb}{0.184314,0.184314,0.184314}%
\pgfsetfillcolor{currentfill}%
\pgfsetfillopacity{0.400000}%
\pgfsetlinewidth{1.505625pt}%
\definecolor{currentstroke}{rgb}{0.184314,0.184314,0.184314}%
\pgfsetstrokecolor{currentstroke}%
\pgfsetstrokeopacity{0.400000}%
\pgfsetdash{}{0pt}%
\pgfsys@defobject{currentmarker}{\pgfqpoint{0.000000in}{-0.026896in}}{\pgfqpoint{0.000000in}{0.026896in}}{%
\pgfpathmoveto{\pgfqpoint{0.000000in}{-0.026896in}}%
\pgfpathlineto{\pgfqpoint{0.000000in}{0.026896in}}%
\pgfusepath{stroke,fill}%
}%
\begin{pgfscope}%
\pgfsys@transformshift{0.385875in}{1.294816in}%
\pgfsys@useobject{currentmarker}{}%
\end{pgfscope}%
\begin{pgfscope}%
\pgfsys@transformshift{0.494375in}{1.294816in}%
\pgfsys@useobject{currentmarker}{}%
\end{pgfscope}%
\begin{pgfscope}%
\pgfsys@transformshift{0.738500in}{1.294816in}%
\pgfsys@useobject{currentmarker}{}%
\end{pgfscope}%
\end{pgfscope}%
\begin{pgfscope}%
\pgfpathrectangle{\pgfqpoint{0.087500in}{0.231000in}}{\pgfqpoint{0.542500in}{1.617000in}}%
\pgfusepath{clip}%
\pgfsetbuttcap%
\pgfsetroundjoin%
\definecolor{currentfill}{rgb}{0.184314,0.184314,0.184314}%
\pgfsetfillcolor{currentfill}%
\pgfsetfillopacity{0.400000}%
\pgfsetlinewidth{1.505625pt}%
\definecolor{currentstroke}{rgb}{0.184314,0.184314,0.184314}%
\pgfsetstrokecolor{currentstroke}%
\pgfsetstrokeopacity{0.400000}%
\pgfsetdash{}{0pt}%
\pgfsys@defobject{currentmarker}{\pgfqpoint{0.000000in}{-0.026896in}}{\pgfqpoint{0.000000in}{0.026896in}}{%
\pgfpathmoveto{\pgfqpoint{0.000000in}{-0.026896in}}%
\pgfpathlineto{\pgfqpoint{0.000000in}{0.026896in}}%
\pgfusepath{stroke,fill}%
}%
\begin{pgfscope}%
\pgfsys@transformshift{0.168875in}{1.379921in}%
\pgfsys@useobject{currentmarker}{}%
\end{pgfscope}%
\begin{pgfscope}%
\pgfsys@transformshift{0.250250in}{1.379921in}%
\pgfsys@useobject{currentmarker}{}%
\end{pgfscope}%
\begin{pgfscope}%
\pgfsys@transformshift{0.467250in}{1.379921in}%
\pgfsys@useobject{currentmarker}{}%
\end{pgfscope}%
\end{pgfscope}%
\begin{pgfscope}%
\pgfpathrectangle{\pgfqpoint{0.087500in}{0.231000in}}{\pgfqpoint{0.542500in}{1.617000in}}%
\pgfusepath{clip}%
\pgfsetbuttcap%
\pgfsetroundjoin%
\definecolor{currentfill}{rgb}{0.184314,0.184314,0.184314}%
\pgfsetfillcolor{currentfill}%
\pgfsetfillopacity{0.400000}%
\pgfsetlinewidth{1.505625pt}%
\definecolor{currentstroke}{rgb}{0.184314,0.184314,0.184314}%
\pgfsetstrokecolor{currentstroke}%
\pgfsetstrokeopacity{0.400000}%
\pgfsetdash{}{0pt}%
\pgfsys@defobject{currentmarker}{\pgfqpoint{0.000000in}{-0.026896in}}{\pgfqpoint{0.000000in}{0.026896in}}{%
\pgfpathmoveto{\pgfqpoint{0.000000in}{-0.026896in}}%
\pgfpathlineto{\pgfqpoint{0.000000in}{0.026896in}}%
\pgfusepath{stroke,fill}%
}%
\begin{pgfscope}%
\pgfsys@transformshift{0.467250in}{1.465026in}%
\pgfsys@useobject{currentmarker}{}%
\end{pgfscope}%
\begin{pgfscope}%
\pgfsys@transformshift{0.684250in}{1.465026in}%
\pgfsys@useobject{currentmarker}{}%
\end{pgfscope}%
\end{pgfscope}%
\begin{pgfscope}%
\pgfpathrectangle{\pgfqpoint{0.087500in}{0.231000in}}{\pgfqpoint{0.542500in}{1.617000in}}%
\pgfusepath{clip}%
\pgfsetbuttcap%
\pgfsetroundjoin%
\definecolor{currentfill}{rgb}{0.184314,0.184314,0.184314}%
\pgfsetfillcolor{currentfill}%
\pgfsetfillopacity{0.400000}%
\pgfsetlinewidth{1.505625pt}%
\definecolor{currentstroke}{rgb}{0.184314,0.184314,0.184314}%
\pgfsetstrokecolor{currentstroke}%
\pgfsetstrokeopacity{0.400000}%
\pgfsetdash{}{0pt}%
\pgfsys@defobject{currentmarker}{\pgfqpoint{0.000000in}{-0.026896in}}{\pgfqpoint{0.000000in}{0.026896in}}{%
\pgfpathmoveto{\pgfqpoint{0.000000in}{-0.026896in}}%
\pgfpathlineto{\pgfqpoint{0.000000in}{0.026896in}}%
\pgfusepath{stroke,fill}%
}%
\begin{pgfscope}%
\pgfsys@transformshift{0.467250in}{1.550132in}%
\pgfsys@useobject{currentmarker}{}%
\end{pgfscope}%
\end{pgfscope}%
\begin{pgfscope}%
\pgfpathrectangle{\pgfqpoint{0.087500in}{0.231000in}}{\pgfqpoint{0.542500in}{1.617000in}}%
\pgfusepath{clip}%
\pgfsetbuttcap%
\pgfsetroundjoin%
\definecolor{currentfill}{rgb}{0.184314,0.184314,0.184314}%
\pgfsetfillcolor{currentfill}%
\pgfsetfillopacity{0.400000}%
\pgfsetlinewidth{1.505625pt}%
\definecolor{currentstroke}{rgb}{0.184314,0.184314,0.184314}%
\pgfsetstrokecolor{currentstroke}%
\pgfsetstrokeopacity{0.400000}%
\pgfsetdash{}{0pt}%
\pgfsys@defobject{currentmarker}{\pgfqpoint{0.000000in}{-0.026896in}}{\pgfqpoint{0.000000in}{0.026896in}}{%
\pgfpathmoveto{\pgfqpoint{0.000000in}{-0.026896in}}%
\pgfpathlineto{\pgfqpoint{0.000000in}{0.026896in}}%
\pgfusepath{stroke,fill}%
}%
\begin{pgfscope}%
\pgfsys@transformshift{0.277375in}{1.635237in}%
\pgfsys@useobject{currentmarker}{}%
\end{pgfscope}%
\begin{pgfscope}%
\pgfsys@transformshift{0.467250in}{1.635237in}%
\pgfsys@useobject{currentmarker}{}%
\end{pgfscope}%
\end{pgfscope}%
\begin{pgfscope}%
\pgfpathrectangle{\pgfqpoint{0.087500in}{0.231000in}}{\pgfqpoint{0.542500in}{1.617000in}}%
\pgfusepath{clip}%
\pgfsetbuttcap%
\pgfsetroundjoin%
\definecolor{currentfill}{rgb}{0.184314,0.184314,0.184314}%
\pgfsetfillcolor{currentfill}%
\pgfsetfillopacity{0.400000}%
\pgfsetlinewidth{1.505625pt}%
\definecolor{currentstroke}{rgb}{0.184314,0.184314,0.184314}%
\pgfsetstrokecolor{currentstroke}%
\pgfsetstrokeopacity{0.400000}%
\pgfsetdash{}{0pt}%
\pgfsys@defobject{currentmarker}{\pgfqpoint{0.000000in}{-0.026896in}}{\pgfqpoint{0.000000in}{0.026896in}}{%
\pgfpathmoveto{\pgfqpoint{0.000000in}{-0.026896in}}%
\pgfpathlineto{\pgfqpoint{0.000000in}{0.026896in}}%
\pgfusepath{stroke,fill}%
}%
\begin{pgfscope}%
\pgfsys@transformshift{0.196000in}{1.720342in}%
\pgfsys@useobject{currentmarker}{}%
\end{pgfscope}%
\begin{pgfscope}%
\pgfsys@transformshift{0.250250in}{1.720342in}%
\pgfsys@useobject{currentmarker}{}%
\end{pgfscope}%
\begin{pgfscope}%
\pgfsys@transformshift{0.331625in}{1.720342in}%
\pgfsys@useobject{currentmarker}{}%
\end{pgfscope}%
\end{pgfscope}%
\begin{pgfscope}%
\pgfpathrectangle{\pgfqpoint{0.087500in}{0.231000in}}{\pgfqpoint{0.542500in}{1.617000in}}%
\pgfusepath{clip}%
\pgfsetbuttcap%
\pgfsetroundjoin%
\definecolor{currentfill}{rgb}{0.184314,0.184314,0.184314}%
\pgfsetfillcolor{currentfill}%
\pgfsetfillopacity{0.400000}%
\pgfsetlinewidth{1.505625pt}%
\definecolor{currentstroke}{rgb}{0.184314,0.184314,0.184314}%
\pgfsetstrokecolor{currentstroke}%
\pgfsetstrokeopacity{0.400000}%
\pgfsetdash{}{0pt}%
\pgfsys@defobject{currentmarker}{\pgfqpoint{0.000000in}{-0.026896in}}{\pgfqpoint{0.000000in}{0.026896in}}{%
\pgfpathmoveto{\pgfqpoint{0.000000in}{-0.026896in}}%
\pgfpathlineto{\pgfqpoint{0.000000in}{0.026896in}}%
\pgfusepath{stroke,fill}%
}%
\end{pgfscope}%
\begin{pgfscope}%
\pgfsetbuttcap%
\pgfsetroundjoin%
\definecolor{currentfill}{rgb}{0.000000,0.000000,0.000000}%
\pgfsetfillcolor{currentfill}%
\pgfsetlinewidth{0.803000pt}%
\definecolor{currentstroke}{rgb}{0.000000,0.000000,0.000000}%
\pgfsetstrokecolor{currentstroke}%
\pgfsetdash{}{0pt}%
\pgfsys@defobject{currentmarker}{\pgfqpoint{0.000000in}{0.000000in}}{\pgfqpoint{0.000000in}{0.048611in}}{%
\pgfpathmoveto{\pgfqpoint{0.000000in}{0.000000in}}%
\pgfpathlineto{\pgfqpoint{0.000000in}{0.048611in}}%
\pgfusepath{stroke,fill}%
}%
\begin{pgfscope}%
\pgfsys@transformshift{0.087500in}{1.848000in}%
\pgfsys@useobject{currentmarker}{}%
\end{pgfscope}%
\end{pgfscope}%
\begin{pgfscope}%
\definecolor{textcolor}{rgb}{0.000000,0.000000,0.000000}%
\pgfsetstrokecolor{textcolor}%
\pgfsetfillcolor{textcolor}%
\pgftext[x=0.087500in,y=1.945222in,,bottom]{\color{textcolor}\rmfamily\fontsize{8.000000}{9.600000}\selectfont 0}%
\end{pgfscope}%
\begin{pgfscope}%
\pgfsetbuttcap%
\pgfsetroundjoin%
\definecolor{currentfill}{rgb}{0.000000,0.000000,0.000000}%
\pgfsetfillcolor{currentfill}%
\pgfsetlinewidth{0.803000pt}%
\definecolor{currentstroke}{rgb}{0.000000,0.000000,0.000000}%
\pgfsetstrokecolor{currentstroke}%
\pgfsetdash{}{0pt}%
\pgfsys@defobject{currentmarker}{\pgfqpoint{0.000000in}{0.000000in}}{\pgfqpoint{0.000000in}{0.048611in}}{%
\pgfpathmoveto{\pgfqpoint{0.000000in}{0.000000in}}%
\pgfpathlineto{\pgfqpoint{0.000000in}{0.048611in}}%
\pgfusepath{stroke,fill}%
}%
\begin{pgfscope}%
\pgfsys@transformshift{0.630000in}{1.848000in}%
\pgfsys@useobject{currentmarker}{}%
\end{pgfscope}%
\end{pgfscope}%
\begin{pgfscope}%
\definecolor{textcolor}{rgb}{0.000000,0.000000,0.000000}%
\pgfsetstrokecolor{textcolor}%
\pgfsetfillcolor{textcolor}%
\pgftext[x=0.630000in,y=1.945222in,,bottom]{\color{textcolor}\rmfamily\fontsize{8.000000}{9.600000}\selectfont 20}%
\end{pgfscope}%
\begin{pgfscope}%
\definecolor{textcolor}{rgb}{0.000000,0.000000,0.000000}%
\pgfsetstrokecolor{textcolor}%
\pgfsetfillcolor{textcolor}%
\pgftext[x=0.358750in,y=1.961190in,,base]{\color{textcolor}\rmfamily\fontsize{8.000000}{9.600000}\selectfont \(\displaystyle \tau\) [ms]}%
\end{pgfscope}%
\begin{pgfscope}%
\pgfsetbuttcap%
\pgfsetmiterjoin%
\definecolor{currentfill}{rgb}{0.000000,0.000000,0.000000}%
\pgfsetfillcolor{currentfill}%
\pgfsetlinewidth{1.003750pt}%
\definecolor{currentstroke}{rgb}{0.000000,0.000000,0.000000}%
\pgfsetstrokecolor{currentstroke}%
\pgfsetdash{}{0pt}%
\pgfsys@defobject{currentmarker}{\pgfqpoint{-0.041667in}{-0.041667in}}{\pgfqpoint{0.041667in}{0.041667in}}{%
\pgfpathmoveto{\pgfqpoint{0.041667in}{-0.000000in}}%
\pgfpathlineto{\pgfqpoint{-0.041667in}{0.041667in}}%
\pgfpathlineto{\pgfqpoint{-0.041667in}{-0.041667in}}%
\pgfpathlineto{\pgfqpoint{0.041667in}{-0.000000in}}%
\pgfpathclose%
\pgfusepath{stroke,fill}%
}%
\begin{pgfscope}%
\pgfsys@transformshift{10.937500in}{1.720342in}%
\pgfsys@useobject{currentmarker}{}%
\end{pgfscope}%
\end{pgfscope}%
\begin{pgfscope}%
\pgfsetrectcap%
\pgfsetmiterjoin%
\pgfsetlinewidth{0.803000pt}%
\definecolor{currentstroke}{rgb}{0.000000,0.000000,0.000000}%
\pgfsetstrokecolor{currentstroke}%
\pgfsetdash{}{0pt}%
\pgfpathmoveto{\pgfqpoint{0.087500in}{1.848000in}}%
\pgfpathlineto{\pgfqpoint{0.630000in}{1.848000in}}%
\pgfusepath{stroke}%
\end{pgfscope}%
\end{pgfpicture}%
\makeatother%
\endgroup%

%% file: example_traces.pgf
\begingroup%
\makeatletter%
\begin{pgfpicture}%
\pgfpathrectangle{\pgfpointorigin}{\pgfqpoint{1.300000in}{0.900000in}}%
\pgfusepath{use as bounding box, clip}%
\begin{pgfscope}%
\pgfsetbuttcap%
\pgfsetmiterjoin%
\pgfsetlinewidth{0.000000pt}%
\definecolor{currentstroke}{rgb}{0.000000,0.000000,0.000000}%
\pgfsetstrokecolor{currentstroke}%
\pgfsetstrokeopacity{0.000000}%
\pgfsetdash{}{0pt}%
\pgfpathmoveto{\pgfqpoint{0.000000in}{0.000000in}}%
\pgfpathlineto{\pgfqpoint{1.300000in}{0.000000in}}%
\pgfpathlineto{\pgfqpoint{1.300000in}{0.900000in}}%
\pgfpathlineto{\pgfqpoint{0.000000in}{0.900000in}}%
\pgfpathlineto{\pgfqpoint{0.000000in}{0.000000in}}%
\pgfpathclose%
\pgfusepath{}%
\end{pgfscope}%
\begin{pgfscope}%
\pgfsetbuttcap%
\pgfsetmiterjoin%
\pgfsetlinewidth{0.000000pt}%
\definecolor{currentstroke}{rgb}{0.000000,0.000000,0.000000}%
\pgfsetstrokecolor{currentstroke}%
\pgfsetstrokeopacity{0.000000}%
\pgfsetdash{}{0pt}%
\pgfpathmoveto{\pgfqpoint{0.325000in}{0.180000in}}%
\pgfpathlineto{\pgfqpoint{1.170000in}{0.180000in}}%
\pgfpathlineto{\pgfqpoint{1.170000in}{0.792000in}}%
\pgfpathlineto{\pgfqpoint{0.325000in}{0.792000in}}%
\pgfpathlineto{\pgfqpoint{0.325000in}{0.180000in}}%
\pgfpathclose%
\pgfusepath{}%
\end{pgfscope}%
\begin{pgfscope}%
\pgfsetbuttcap%
\pgfsetroundjoin%
\definecolor{currentfill}{rgb}{0.000000,0.000000,0.000000}%
\pgfsetfillcolor{currentfill}%
\pgfsetlinewidth{0.803000pt}%
\definecolor{currentstroke}{rgb}{0.000000,0.000000,0.000000}%
\pgfsetstrokecolor{currentstroke}%
\pgfsetdash{}{0pt}%
\pgfsys@defobject{currentmarker}{\pgfqpoint{0.000000in}{-0.048611in}}{\pgfqpoint{0.000000in}{0.000000in}}{%
\pgfpathmoveto{\pgfqpoint{0.000000in}{0.000000in}}%
\pgfpathlineto{\pgfqpoint{0.000000in}{-0.048611in}}%
\pgfusepath{stroke,fill}%
}%
\begin{pgfscope}%
\pgfsys@transformshift{0.325000in}{0.180000in}%
\pgfsys@useobject{currentmarker}{}%
\end{pgfscope}%
\end{pgfscope}%
\begin{pgfscope}%
\definecolor{textcolor}{rgb}{0.000000,0.000000,0.000000}%
\pgfsetstrokecolor{textcolor}%
\pgfsetfillcolor{textcolor}%
\pgftext[x=0.325000in,y=0.082778in,,top]{\color{textcolor}\rmfamily\fontsize{6.000000}{7.200000}\selectfont 0}%
\end{pgfscope}%
\begin{pgfscope}%
\pgfsetbuttcap%
\pgfsetroundjoin%
\definecolor{currentfill}{rgb}{0.000000,0.000000,0.000000}%
\pgfsetfillcolor{currentfill}%
\pgfsetlinewidth{0.803000pt}%
\definecolor{currentstroke}{rgb}{0.000000,0.000000,0.000000}%
\pgfsetstrokecolor{currentstroke}%
\pgfsetdash{}{0pt}%
\pgfsys@defobject{currentmarker}{\pgfqpoint{0.000000in}{-0.048611in}}{\pgfqpoint{0.000000in}{0.000000in}}{%
\pgfpathmoveto{\pgfqpoint{0.000000in}{0.000000in}}%
\pgfpathlineto{\pgfqpoint{0.000000in}{-0.048611in}}%
\pgfusepath{stroke,fill}%
}%
\begin{pgfscope}%
\pgfsys@transformshift{1.170000in}{0.180000in}%
\pgfsys@useobject{currentmarker}{}%
\end{pgfscope}%
\end{pgfscope}%
\begin{pgfscope}%
\definecolor{textcolor}{rgb}{0.000000,0.000000,0.000000}%
\pgfsetstrokecolor{textcolor}%
\pgfsetfillcolor{textcolor}%
\pgftext[x=1.170000in,y=0.082778in,,top]{\color{textcolor}\rmfamily\fontsize{6.000000}{7.200000}\selectfont 25}%
\end{pgfscope}%
\begin{pgfscope}%
\definecolor{textcolor}{rgb}{0.000000,0.000000,0.000000}%
\pgfsetstrokecolor{textcolor}%
\pgfsetfillcolor{textcolor}%
\pgftext[x=0.747500in,y=0.118800in,,top]{\color{textcolor}\rmfamily\fontsize{6.000000}{7.200000}\selectfont \(\displaystyle \tau\) [ms]}%
\end{pgfscope}%
\begin{pgfscope}%
\pgfsetbuttcap%
\pgfsetroundjoin%
\definecolor{currentfill}{rgb}{0.000000,0.000000,0.000000}%
\pgfsetfillcolor{currentfill}%
\pgfsetlinewidth{0.803000pt}%
\definecolor{currentstroke}{rgb}{0.000000,0.000000,0.000000}%
\pgfsetstrokecolor{currentstroke}%
\pgfsetdash{}{0pt}%
\pgfsys@defobject{currentmarker}{\pgfqpoint{-0.048611in}{0.000000in}}{\pgfqpoint{-0.000000in}{0.000000in}}{%
\pgfpathmoveto{\pgfqpoint{-0.000000in}{0.000000in}}%
\pgfpathlineto{\pgfqpoint{-0.048611in}{0.000000in}}%
\pgfusepath{stroke,fill}%
}%
\begin{pgfscope}%
\pgfsys@transformshift{0.325000in}{0.216298in}%
\pgfsys@useobject{currentmarker}{}%
\end{pgfscope}%
\end{pgfscope}%
\begin{pgfscope}%
\definecolor{textcolor}{rgb}{0.000000,0.000000,0.000000}%
\pgfsetstrokecolor{textcolor}%
\pgfsetfillcolor{textcolor}%
\pgftext[x=0.015004in, y=0.184641in, left, base]{\color{textcolor}\rmfamily\fontsize{6.000000}{7.200000}\selectfont \ensuremath{-}0.5}%
\end{pgfscope}%
\begin{pgfscope}%
\pgfsetbuttcap%
\pgfsetroundjoin%
\definecolor{currentfill}{rgb}{0.000000,0.000000,0.000000}%
\pgfsetfillcolor{currentfill}%
\pgfsetlinewidth{0.803000pt}%
\definecolor{currentstroke}{rgb}{0.000000,0.000000,0.000000}%
\pgfsetstrokecolor{currentstroke}%
\pgfsetdash{}{0pt}%
\pgfsys@defobject{currentmarker}{\pgfqpoint{-0.048611in}{0.000000in}}{\pgfqpoint{-0.000000in}{0.000000in}}{%
\pgfpathmoveto{\pgfqpoint{-0.000000in}{0.000000in}}%
\pgfpathlineto{\pgfqpoint{-0.048611in}{0.000000in}}%
\pgfusepath{stroke,fill}%
}%
\begin{pgfscope}%
\pgfsys@transformshift{0.325000in}{0.727709in}%
\pgfsys@useobject{currentmarker}{}%
\end{pgfscope}%
\end{pgfscope}%
\begin{pgfscope}%
\definecolor{textcolor}{rgb}{0.000000,0.000000,0.000000}%
\pgfsetstrokecolor{textcolor}%
\pgfsetfillcolor{textcolor}%
\pgftext[x=0.095250in, y=0.696052in, left, base]{\color{textcolor}\rmfamily\fontsize{6.000000}{7.200000}\selectfont 1.0}%
\end{pgfscope}%
\begin{pgfscope}%
\definecolor{textcolor}{rgb}{0.000000,0.000000,0.000000}%
\pgfsetstrokecolor{textcolor}%
\pgfsetfillcolor{textcolor}%
\pgftext[x=0.240500in,y=0.486000in,,bottom,rotate=90.000000]{\color{textcolor}\rmfamily\fontsize{6.000000}{7.200000}\selectfont \(\displaystyle v_k\) [a.u.]}%
\end{pgfscope}%
\begin{pgfscope}%
\pgfpathrectangle{\pgfqpoint{0.325000in}{0.180000in}}{\pgfqpoint{0.845000in}{0.612000in}}%
\pgfusepath{clip}%
\pgfsetrectcap%
\pgfsetroundjoin%
\pgfsetlinewidth{1.003750pt}%
\definecolor{currentstroke}{rgb}{1.000000,0.000000,0.000000}%
\pgfsetstrokecolor{currentstroke}%
\pgfsetdash{}{0pt}%
\pgfpathmoveto{\pgfqpoint{0.325000in}{0.386768in}}%
\pgfpathlineto{\pgfqpoint{0.358800in}{0.386768in}}%
\pgfpathlineto{\pgfqpoint{0.392600in}{0.386768in}}%
\pgfpathlineto{\pgfqpoint{0.426400in}{0.372153in}}%
\pgfpathlineto{\pgfqpoint{0.460200in}{0.346073in}}%
\pgfpathlineto{\pgfqpoint{0.494000in}{0.315481in}}%
\pgfpathlineto{\pgfqpoint{0.527800in}{0.272270in}}%
\pgfpathlineto{\pgfqpoint{0.561600in}{0.248918in}}%
\pgfpathlineto{\pgfqpoint{0.595400in}{0.236530in}}%
\pgfpathlineto{\pgfqpoint{0.629200in}{0.221615in}}%
\pgfpathlineto{\pgfqpoint{0.663000in}{0.207818in}}%
\pgfpathlineto{\pgfqpoint{0.696800in}{0.212293in}}%
\pgfpathlineto{\pgfqpoint{0.730600in}{0.220247in}}%
\pgfpathlineto{\pgfqpoint{0.764400in}{0.230397in}}%
\pgfpathlineto{\pgfqpoint{0.798200in}{0.254676in}}%
\pgfpathlineto{\pgfqpoint{0.832000in}{0.284291in}}%
\pgfpathlineto{\pgfqpoint{0.865800in}{0.307704in}}%
\pgfpathlineto{\pgfqpoint{0.899600in}{0.326159in}}%
\pgfpathlineto{\pgfqpoint{0.933400in}{0.340658in}}%
\pgfpathlineto{\pgfqpoint{0.967200in}{0.352008in}}%
\pgfpathlineto{\pgfqpoint{1.001000in}{0.360856in}}%
\pgfpathlineto{\pgfqpoint{1.034800in}{0.367719in}}%
\pgfpathlineto{\pgfqpoint{1.068600in}{0.381597in}}%
\pgfpathlineto{\pgfqpoint{1.102400in}{0.391377in}}%
\pgfpathlineto{\pgfqpoint{1.136200in}{0.415521in}}%
\pgfpathlineto{\pgfqpoint{1.170000in}{0.431489in}}%
\pgfpathlineto{\pgfqpoint{1.175000in}{0.432946in}}%
\pgfusepath{stroke}%
\end{pgfscope}%
\begin{pgfscope}%
\pgfpathrectangle{\pgfqpoint{0.325000in}{0.180000in}}{\pgfqpoint{0.845000in}{0.612000in}}%
\pgfusepath{clip}%
\pgfsetrectcap%
\pgfsetroundjoin%
\pgfsetlinewidth{1.003750pt}%
\definecolor{currentstroke}{rgb}{0.000000,0.000000,1.000000}%
\pgfsetstrokecolor{currentstroke}%
\pgfsetdash{}{0pt}%
\pgfpathmoveto{\pgfqpoint{0.325000in}{0.386768in}}%
\pgfpathlineto{\pgfqpoint{0.358800in}{0.386768in}}%
\pgfpathlineto{\pgfqpoint{0.392600in}{0.386768in}}%
\pgfpathlineto{\pgfqpoint{0.426400in}{0.396457in}}%
\pgfpathlineto{\pgfqpoint{0.460200in}{0.393647in}}%
\pgfpathlineto{\pgfqpoint{0.494000in}{0.400378in}}%
\pgfpathlineto{\pgfqpoint{0.527800in}{0.402815in}}%
\pgfpathlineto{\pgfqpoint{0.561600in}{0.390605in}}%
\pgfpathlineto{\pgfqpoint{0.595400in}{0.382019in}}%
\pgfpathlineto{\pgfqpoint{0.629200in}{0.372683in}}%
\pgfpathlineto{\pgfqpoint{0.663000in}{0.355227in}}%
\pgfpathlineto{\pgfqpoint{0.696800in}{0.346385in}}%
\pgfpathlineto{\pgfqpoint{0.730600in}{0.341367in}}%
\pgfpathlineto{\pgfqpoint{0.764400in}{0.339143in}}%
\pgfpathlineto{\pgfqpoint{0.798200in}{0.346760in}}%
\pgfpathlineto{\pgfqpoint{0.832000in}{0.355566in}}%
\pgfpathlineto{\pgfqpoint{0.865800in}{0.362548in}}%
\pgfpathlineto{\pgfqpoint{0.899600in}{0.368069in}}%
\pgfpathlineto{\pgfqpoint{0.933400in}{0.372422in}}%
\pgfpathlineto{\pgfqpoint{0.967200in}{0.375844in}}%
\pgfpathlineto{\pgfqpoint{1.001000in}{0.378524in}}%
\pgfpathlineto{\pgfqpoint{1.034800in}{0.380614in}}%
\pgfpathlineto{\pgfqpoint{1.068600in}{0.387599in}}%
\pgfpathlineto{\pgfqpoint{1.102400in}{0.392427in}}%
\pgfpathlineto{\pgfqpoint{1.136200in}{0.403080in}}%
\pgfpathlineto{\pgfqpoint{1.170000in}{0.410026in}}%
\pgfpathlineto{\pgfqpoint{1.175000in}{0.410644in}}%
\pgfusepath{stroke}%
\end{pgfscope}%
\begin{pgfscope}%
\pgfpathrectangle{\pgfqpoint{0.325000in}{0.180000in}}{\pgfqpoint{0.845000in}{0.612000in}}%
\pgfusepath{clip}%
\pgfsetrectcap%
\pgfsetroundjoin%
\pgfsetlinewidth{1.003750pt}%
\definecolor{currentstroke}{rgb}{1.000000,0.647059,0.000000}%
\pgfsetstrokecolor{currentstroke}%
\pgfsetdash{}{0pt}%
\pgfpathmoveto{\pgfqpoint{0.325000in}{0.386768in}}%
\pgfpathlineto{\pgfqpoint{0.358800in}{0.386768in}}%
\pgfpathlineto{\pgfqpoint{0.392600in}{0.386768in}}%
\pgfpathlineto{\pgfqpoint{0.426400in}{0.394263in}}%
\pgfpathlineto{\pgfqpoint{0.460200in}{0.411820in}}%
\pgfpathlineto{\pgfqpoint{0.494000in}{0.424778in}}%
\pgfpathlineto{\pgfqpoint{0.527800in}{0.450199in}}%
\pgfpathlineto{\pgfqpoint{0.561600in}{0.475047in}}%
\pgfpathlineto{\pgfqpoint{0.595400in}{0.489850in}}%
\pgfpathlineto{\pgfqpoint{0.629200in}{0.503108in}}%
\pgfpathlineto{\pgfqpoint{0.663000in}{0.522597in}}%
\pgfpathlineto{\pgfqpoint{0.696800in}{0.535796in}}%
\pgfpathlineto{\pgfqpoint{0.730600in}{0.540821in}}%
\pgfpathlineto{\pgfqpoint{0.764400in}{0.540032in}}%
\pgfpathlineto{\pgfqpoint{0.798200in}{0.566565in}}%
\pgfpathlineto{\pgfqpoint{0.832000in}{0.583433in}}%
\pgfpathlineto{\pgfqpoint{0.865800in}{0.589685in}}%
\pgfpathlineto{\pgfqpoint{0.899600in}{0.588389in}}%
\pgfpathlineto{\pgfqpoint{0.933400in}{0.581889in}}%
\pgfpathlineto{\pgfqpoint{0.967200in}{0.571955in}}%
\pgfpathlineto{\pgfqpoint{1.001000in}{0.559912in}}%
\pgfpathlineto{\pgfqpoint{1.034800in}{0.546740in}}%
\pgfpathlineto{\pgfqpoint{1.068600in}{0.536991in}}%
\pgfpathlineto{\pgfqpoint{1.102400in}{0.526049in}}%
\pgfpathlineto{\pgfqpoint{1.136200in}{0.510080in}}%
\pgfpathlineto{\pgfqpoint{1.170000in}{0.495566in}}%
\pgfpathlineto{\pgfqpoint{1.175000in}{0.493628in}}%
\pgfusepath{stroke}%
\end{pgfscope}%
\begin{pgfscope}%
\pgfpathrectangle{\pgfqpoint{0.325000in}{0.180000in}}{\pgfqpoint{0.845000in}{0.612000in}}%
\pgfusepath{clip}%
\pgfsetrectcap%
\pgfsetroundjoin%
\pgfsetlinewidth{2.007500pt}%
\definecolor{currentstroke}{rgb}{0.000000,0.501961,0.000000}%
\pgfsetstrokecolor{currentstroke}%
\pgfsetdash{}{0pt}%
\pgfpathmoveto{\pgfqpoint{0.325000in}{0.386768in}}%
\pgfpathlineto{\pgfqpoint{0.358800in}{0.386768in}}%
\pgfpathlineto{\pgfqpoint{0.392600in}{0.386768in}}%
\pgfpathlineto{\pgfqpoint{0.426400in}{0.401487in}}%
\pgfpathlineto{\pgfqpoint{0.460200in}{0.441851in}}%
\pgfpathlineto{\pgfqpoint{0.494000in}{0.499073in}}%
\pgfpathlineto{\pgfqpoint{0.527800in}{0.572934in}}%
\pgfpathlineto{\pgfqpoint{0.561600in}{0.636394in}}%
\pgfpathlineto{\pgfqpoint{0.595400in}{0.673529in}}%
\pgfpathlineto{\pgfqpoint{0.629200in}{0.712211in}}%
\pgfpathlineto{\pgfqpoint{0.663000in}{0.754572in}}%
\pgfpathlineto{\pgfqpoint{0.696800in}{0.764182in}}%
\pgfpathlineto{\pgfqpoint{0.730600in}{0.760371in}}%
\pgfpathlineto{\pgfqpoint{0.764400in}{0.747348in}}%
\pgfpathlineto{\pgfqpoint{0.798200in}{0.700883in}}%
\pgfpathlineto{\pgfqpoint{0.832000in}{0.650300in}}%
\pgfpathlineto{\pgfqpoint{0.865800in}{0.607853in}}%
\pgfpathlineto{\pgfqpoint{0.899600in}{0.572234in}}%
\pgfpathlineto{\pgfqpoint{0.933400in}{0.542347in}}%
\pgfpathlineto{\pgfqpoint{0.967200in}{0.517271in}}%
\pgfpathlineto{\pgfqpoint{1.001000in}{0.496231in}}%
\pgfpathlineto{\pgfqpoint{1.034800in}{0.478580in}}%
\pgfpathlineto{\pgfqpoint{1.068600in}{0.453365in}}%
\pgfpathlineto{\pgfqpoint{1.102400in}{0.434004in}}%
\pgfpathlineto{\pgfqpoint{1.136200in}{0.405671in}}%
\pgfpathlineto{\pgfqpoint{1.170000in}{0.385470in}}%
\pgfpathlineto{\pgfqpoint{1.175000in}{0.383400in}}%
\pgfusepath{stroke}%
\end{pgfscope}%
\begin{pgfscope}%
\pgfsetrectcap%
\pgfsetmiterjoin%
\pgfsetlinewidth{0.803000pt}%
\definecolor{currentstroke}{rgb}{0.000000,0.000000,0.000000}%
\pgfsetstrokecolor{currentstroke}%
\pgfsetdash{}{0pt}%
\pgfpathmoveto{\pgfqpoint{0.325000in}{0.180000in}}%
\pgfpathlineto{\pgfqpoint{0.325000in}{0.792000in}}%
\pgfusepath{stroke}%
\end{pgfscope}%
\begin{pgfscope}%
\pgfsetrectcap%
\pgfsetmiterjoin%
\pgfsetlinewidth{0.803000pt}%
\definecolor{currentstroke}{rgb}{0.000000,0.000000,0.000000}%
\pgfsetstrokecolor{currentstroke}%
\pgfsetdash{}{0pt}%
\pgfpathmoveto{\pgfqpoint{0.325000in}{0.180000in}}%
\pgfpathlineto{\pgfqpoint{1.170000in}{0.180000in}}%
\pgfusepath{stroke}%
\end{pgfscope}%
\end{pgfpicture}%
\makeatother%
\endgroup%

%% file: results.tex
\section{Results and Conclusions}

\begin{figure}[tb]
    \centering
    \footnotesize
    \tikzset{
        panel/.style={
            inner sep=0pt, outer sep=0, execute at begin node={\tikzset{anchor=center, inner sep=.33333em}}},
        label/.style={anchor=north west, inner sep=0, outer sep=0}}
    \begin{tikzpicture}

    \node[panel, anchor=north west] (a) at (0,  2) {
        \input{imdd_model.tex}};
    \node[label] at (a.north west) {\textbf{A}};
    
    \node[panel, anchor=north west] (b) at (0.2, -0.3) {
        \input{params.tex}};
    \node[label] at (0., 0.) {\textbf{B}};
    
    \node[panel, anchor=north west] (c) at (4., 0.) {
        \input{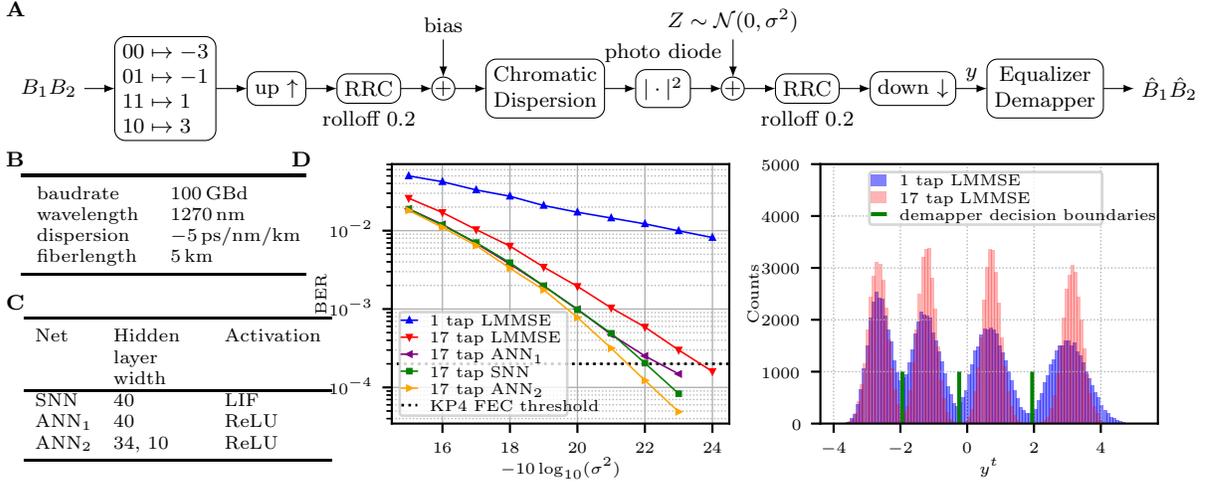}};
    \node[label] at (3.8, 0.0) {\textbf{D}};
    
    \node[panel, anchor=north west] (d) at (0.15, -2.2) {
        \input{struct.tex}};
    \node[label] at (0.0, -1.9) {\textbf{C}};

    \end{tikzpicture}
    
    \caption{\footnotesize
    \textbf{(A)} Simulated IM/DD link. \textbf{(B)} IM/DD parameters. \textbf{(C)} NN equalizer parameters. \textbf{(D)} Left: BER results for transmission of PAM4
over the simulated IM/DD link. Right: Histogram of the linear MMSE equalizer output. }
    \vspace{-8pt}
    \label{fig:models}
\end{figure}

In Fig.~\ref{fig:models}A, we display a simulated IM/DD link. Bits are mapped to a PAM4 constellation, the signal is upsampled and filtered by a \emph{root-raised-cosine} (RRC). The signal is then shifted to the positive and CD is applied. At the receiver, a PD squares the signal and AWGN is added. The signal is then RRC filtered and downsampled. The resulting signal $y$ is equalized and demapped. As reference, we use a digital 17 tap LMMSE equalizer, followed by a demapper with BER optimized decision boundaries, see Fig.~\ref{fig:models}D (right), \tcb{and ANNs with one and two hidden layers, respectively, see Fig.~\ref{fig:models}C}.
In Fig.~\ref{fig:models}D (left) we see that joint equalization and demapping by a 17 tap SNN \tcb{outperforms the LMMSE, and performs as well as the 17 tap ANN$_1$, which has 1 hidden layer with 40 neurons, similar to the SNN}. The reference schemes and the SNN were trained using supervised learning.

%% file: imdd_model.tex
\begin{tikzpicture}
    \tikzset{device/.style={align=center, draw, rounded corners}}
    \matrix[column sep=0.4cm, ampersand replacement=\&]{
    \node(src){$B_1B_2$};
    \& \node[device,align=left](map){$00\mapsto -3$\\
    $01\mapsto -1$\\
    $11\mapsto 1$\\
    $10\mapsto 3$};
    \& \node[device](up){up $\uparrow$};
    \& \node[device](rrc){RRC};
    \& \node[circle, draw, inner sep=0cm](bias){$+$};
    \& \node[device](cd){Chromatic\\Dispersion};
    \& \node[device](pd){$|\cdot|^2$};
    \& \node[circle, draw, inner sep=0cm](noise){$+$};
    \& \node[device](rrc_rx){RRC};
    \& \node[device](down){down $\downarrow$};
    \& \node[device](demap){Equalizer\\ Demapper};
    \& \node(sink){$\hat{B}_1\hat{B}_2$};\\
    };
    \draw[-latex](src)--(map);
    \draw[-latex](map)--(up);
    \draw[-latex](up)--(rrc);
    \draw[-latex](rrc)--(bias);
    \draw[-latex](bias)--(cd);
    \draw[-latex](cd)--(pd);
    \draw[-latex](pd)--(noise);
    \draw[-latex](noise)--(rrc_rx);
    \draw[-latex](rrc_rx)--(down);
    \draw[-latex](down)--node[above]{$y$}(demap);
    \draw[-latex](demap)--(sink);
    
    \node[above=0.5cm of bias](bias_val){bias};
    \draw[-latex](bias_val)--(bias);
    
    \node[above=0.5cm of noise](noise_val){$Z\sim\mathcal{N}(0,\sigma^2)$};
    \draw[-latex](noise_val)--(noise);
    
    \node[below=0cm of rrc]{rolloff $0.2$};
    \node[below=0cm of rrc_rx]{rolloff $0.2$};
    
    \node[above=0cm of pd]{photo diode};
\end{tikzpicture}

%% file: params.tex

\scriptsize
\begin{tabular}{ll}\toprule
    baudrate & \SI{100}{GBd} \\
    wavelength & \SI{1270}{nm} \\
    dispersion & \SI{-5}{ps/nm/km} \\
    fiberlength & \SI{5}{km}\\\bottomrule
\end{tabular}

%% file: struct.tex
\setlength{\tabcolsep}{4pt}
\scriptsize
\begin{tabular}{lp{1.2cm}l}\toprule
    \scriptsize Net & \scriptsize Hidden layer width & \scriptsize Activation \\ \hline
    SNN & 40 & LIF \\
    ANN$_1$ & 40 & ReLU \\
    ANN$_2$ & 34, 10 & ReLU \\\bottomrule
\end{tabular}

%
%
%

%% file: discussion.tex
By means of software simulation, we have shown that an SNN suitable for analog electronic hardware can \tcb{efficiently compensate impairments in a simulated IM/DD link}.
In ongoing research, we implement the proposed SNN on the BSS-2 system, with the aim to reproduce the reported results on analog hardware.